\documentclass[]{spie}  

\usepackage{amsmath,amsfonts,amssymb}
\usepackage{graphicx}
\usepackage[colorlinks=true, allcolors=blue]{hyperref}
\usepackage{siunitx}
\usepackage[]{xspace}
\usepackage{enumitem}

\newcommand{\arcsec}{\mbox{$''$}}

\newcommand{\halpha}{\mbox{ $\mathrm{H\alpha}$}\xspace}
\newcommand{\hbeta}{\mbox{ $\mathrm{H\beta}$}\xspace}
\newcommand{\mjup}{\mbox{ $\mathrm{M_{Jup}}$}\xspace}
\newcommand{\mjupy}{\mbox{ $\mathrm{M_{Jup} yr^{-1}}$}\xspace}

\newcommand{\msuny}{\mbox{ $\mathrm{M_{Sun} yr^{-1}}$}\xspace}

\newcommand{\micron}{$\SI{}{\, \micro\meter}$\xspace}

\def\begini{\begin{itemize}[itemsep=0.5pt,topsep=0.5pt]}
\def\endi{\end{itemize}}

\title{Original use of MUSE's laser tomography adaptive optics \\ to directly image young accreting exoplanets}

\author[a]{Julien H. Girard}
\author[b]{Jozua de Boer}
\author[b,c]{Sebastiaan Haffert}
\author[d, a]{Peter Zeidler}
\author[b]{Alexander Bohn}
\author[b]{Rob G. van Holstein}
\author[b]{Ignas Snellen}
\author[b,e]{Jarle Brinchmann}
\author[b]{Christoph Keller}
\author[f]{Roland Bacon}
\author[g]{Jaehan Bae}

\affil[a]{Space Telescope Science Institute, 3700 San Martin Dr, Baltimore MD, 21218, USA}
\affil[b]{Leiden Observatory, Leiden University, PO Box 9513, 2300 RA, Leiden, The Netherlands}
\affil[c]{Astronomy Department, University of Arizona, 933 N. Cherry Ave., Tucson, AZ 85721, USA}
\affil[d]{Department of Physics and Astronomy, Johns Hopkins University, Baltimore, MD 21218, USA}
\affil[e]{Instituto de Astrof\`{i}sica e C\^{i}encias do Espa\c{c}o, Universidade do Porto, CAUP, Rua das Estrelas, PT4150-762 Porto, Portugal}
\affil[f]{Univ Lyon, Univ Lyon1, Ens de Lyon, CNRS, Centre de Recherche Astrophysique de Lyon UMR5574, F-69230, Saint-Genis-Laval, France}
\affil[g]{Department of Terrestrial Magnetism, Carnegie Institution for Science, 5241 Broad Branch Road,
NW, Washington, DC 20015, USA}

\begin{document} 
\maketitle

\begin{abstract}
We present recent results obtained with the VLT/MUSE Integral Field Spectrograph\cite{bacon2014} fed by the 4LGSF\cite{madec2016} and its laser tomography adaptive optics module GALACSI\cite{lapenna2016}. While this so-called narrow-field mode of MUSE was not designed to perform directly imaging of exoplanets and outflows, we show that it can be a game changer to detect and characterize young exoplanets with a prominent emission lines (i.e  \halpha, tracer of accretion), at moderate contrasts. These performances are achieved thanks to the combo of a near-diffraction limited PSF and a medium resolution spectrograph and a cross-correlation approach in post-processing . We discuss this in the context of ground and space, infrared and visible wavelengths, preparing for missions like JWST and WFIRST in great synergy and as pathfinder for future ELT/GSMT (Extremely Large and/or Giant Segmented Mirror Telescopes) instruments\cite{hippler2019}.

\end{abstract}

\keywords{VLT MUSE, laser tomography adaptive optics, high contrast imaging, exoplanets, integral field spectroscopy, planet formation}

\section{INTRODUCTION}
\label{sec:intro}  

Unraveling the keys of planet formation is one great piece of the puzzle that is our quest to understand how our solar system was formed and its singularity or normality with respect to numerous other exoplanetary systems in our galaxy and throughout the Universe. As a community, we have an urgency to understand the links between star and planet formation, the planet-disk interactions, the demographics, accretion, evolution and migration processes of all planets and other constituents of the circumstellar material. There are currently several formation scenarios for planets and a "fauna" of several thousand exoplanets detected with several detection techniques to feed models, confirm or constraint theories. The three main indirect methods - radial velocity, transit, and microlensing - provided most detections and already answered a lot of questions: exoplanets are frequent, they may even outnumber the stars. Small "Earth-like" planets are more frequent than giant gaseous planets. Exoplanetary systems are very diverse.

\subsection{A golden age for direct imaging of star and planet formation sites}

Direct imaging (DI) opens the parameter space to planets at larger physical separations, from 5 to a few 100 of astronomical units (au). So far DI has been successful with HST and from the ground with 6.5 to 10m telescopes operating with adaptive optics in the near to mid IR. Those wavelengths are favorable to image young/hot, self-luminous jupiter analogs which are only a few hundreds to a few tens of thousands fainter than their host star, depending on how young the system is. Nevertheless, the largest surveys so far (GPIES, Nielsen et al. 2019\cite{nielsen2019} and SPHERE/SHINE, Vigan et al. {\it in prep}) of hundreds of young nearby stars tell us - with statistical significance - that giant gaseous planets at large separation like the quatuor around HR8799\cite{marois2008, marois2010_hr8799e} are rare. It is non-trivial to converge on which formation scenarii are more likely and questions remain mostly unanswered:  
\begini
 \item Were these planets formed close to the star or more "in situ", further out in the disk?
 \item Was migration involved? Both ways?
\endi

\subsection{ALMA, xAO: a lot of disks, gaps}

While the two most productive extreme Adaptive Optics (xAO) systems SPHERE and GPI have only detected 3 nw planets, a huge number of disks have been imaged since they started operations in 2013-2014.

\begin{figure}[h!]
\begin{center}
\includegraphics[width=0.75\textwidth, angle=0]{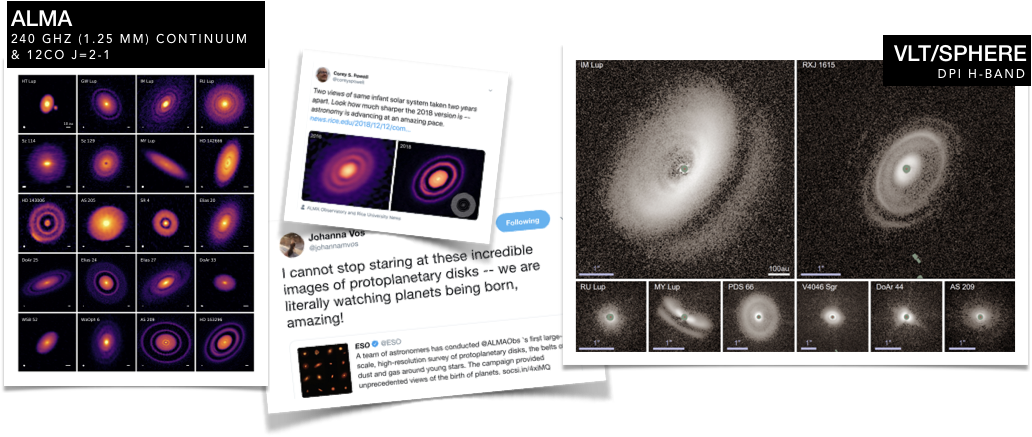}
\caption{Left: ALMA 240 GHz continuum images from the DSHARP\cite{andrews2018} program that left the community and the public in total awe (Middle, with a couple tweets). Right: The same thing is true for the TTauri disks imaged in scattered light by VLT/SPHERE/IRDIS\cite{deboer2019dpi, vanholstein2019dpi} from the DARTTS-S\cite{avenhaus2018} program.}
\label{fig:where}
\end{center}
\end{figure}

Figure~\ref{fig:where} shows that both ALMA and xAO  have revealed young circumstellar disks, planet formation sites, with a great deal of sharpness and contrast, showing evidence of multipole rings and gaps in young systems. These images have similar spatial resolutions (10 to 100 mas) and that shows the tremendous synergy from such facilities. Sub-mm observation can trace cold gas and dust while near-IR in polarized scattered light can reveal the distribution of warmer dust. 

\begini
 \item Where are the planets?
\item How are gas-giant planets and smaller assembled?
\item What is their distribution of accretion rates and entropies?
\item How steady or stochastic is planetary accretion?
\endi

Garufi et al. 2018\cite{garufi2018} attempted a "taxonomy" of protoplanetary disks and showed possible trends. Disks of younger systems are faint, spirals take time to form, ring disks are very common. Transition disks around intermediate to massive stars have large cavities and seem to evolve on larger time scales than those around lower-mass hosts. Several studies have determined the presence of cold giant planets in systems like, for instance, HD 163296\cite{pinte2019}. Yet, many times either current facilities do not probe the right parameter space, either detections are vain or ambiguous due to the presence of relatively bright complex disk structures that can mimic or outshine point-source like planets.

\section{PDS 70, a new benchmark system}

PDS 70 is a K5 type, weak emission TTauri star of $\sim$0.85 solar mass and $\sim$5-6 Myr age. It is located in the Upper Centaurus Lupus moving group at 113 pc distance. It has a prominent ring-like gas/dust (transition) disk with an outer semi-major axis of $\sim$140 au and harbors a giant and sharp,$\sim$70 au clear cavity inside. The PDS 70 disk is an ideal site to look for disk-planet(s) interaction. Radial segregation of dust grains as observed with ALMA is possibly generated by a radial pressure gradient in the disk. A potential mechanism is the dynamical clearing of the disk by a planetary-mass companion\cite{pinilla2013}. 

\subsection{PDS 70 from 2005 to 2012}

With VLT/NACO four-quadrant phase mask coronagraphy at moderate Streh ratios ($\leq$ 10-20\%)  in the J-band, Riaud et al. 2006 (\cite{riaud2006} showed a marginally detected disk with the correct position angle. Later, in 2012, Hashimoto et al. \cite{hashimoto2012} published a Gemini/NICI image in L'-band displaying the disk and the best image thus far using the Subaru/HiCiAO instrument in differential polarimetry. The latter showed a nice ring-link disk ("rim category in the Garufi classification) with a good understanding of the location of near and far sides and a clear gap. The outer disk radius of about 140 au and for the first time, a surface brightness deficit hints a possible disk-planet interaction.

\subsection{The discovery and confirmations of planet b}

And indeed, in 2018, Keppler et al. 2018\cite{keppler2018} \& M\"{u}ller et al. 2018\cite{muller2018} claimed the robust detection of a $\sim$4-17 \mjup bound protoplanet well in the disk gap at $\sim$0.2\arcsec separation, the later one with great signal to noise ratio. Both the companion and the disk were imaged simultaneously in the NIR with VLT/SPHERE, NACO and suing the Gemini/NICI archive (one could guess its detection in the 2012 paper but the data had to be reprocessed). Shortly after, Wagner et al. 2018 claimed a low SNR detection of PDS 70 b in \halpha (656 nm) with MagAO using narrow-band spectral differential imaging (NBSDI). PDS 70 b is co-moving, accreting and responsible of at least some of the carved gap. 

\section{MUSE narrow-field mode (NFM)}

MUSE NFM came as the "new kid on the block", unexpectedly performant. Our team (Julien Girard et al. for a Science Verification project described in figure~\ref{fig:svgirard}, Sebastiaan Haffert for a the commissioning run) decided to capitalize on previous results with long-slit and IFU spectrographs where the gain in contrast was brought by both AO and a somewhat high (R $\geq$ 1,500) spectral resolution. In figure~\ref{fig:parameterspace} one can see the type of PSF one gets from MUSE/NFM, a interesting parameter/discovery space as shown on figure~\ref{fig:parameterspace} assuming putative protoplanets accrete a lot and their young host stars not so much.

\begin{figure}[h!]
\begin{center}
\includegraphics[width=0.85\textwidth, angle=0]{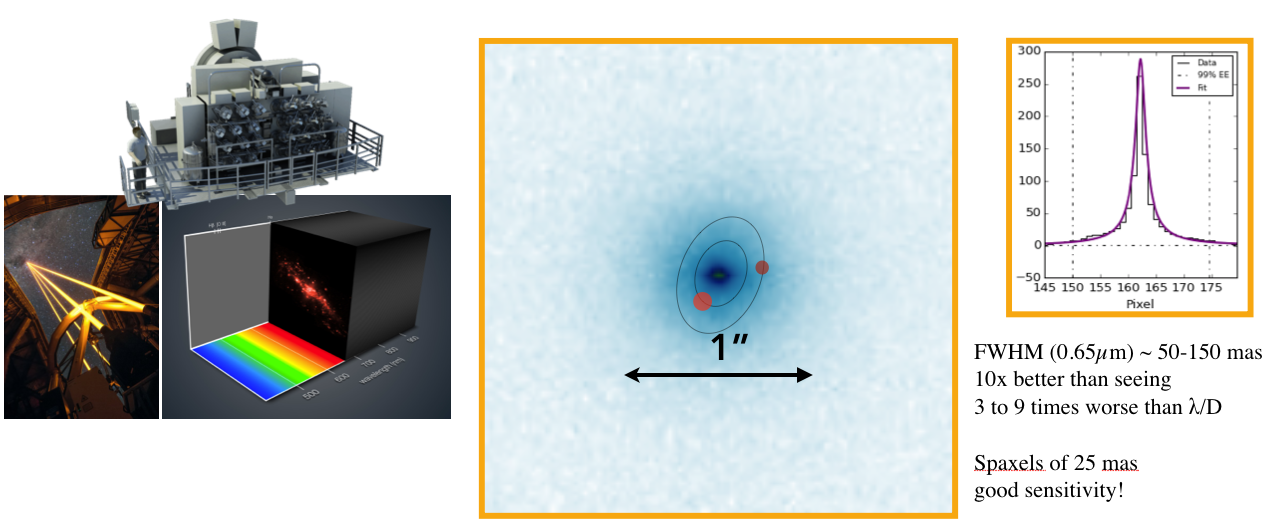}
\caption{Left: The MUSE instrument with its 24 integral field spectrograph units (IFUs) is attached to VLT/UT4. 4 lasers can feed 4 wavefront sensors. The wavefront correction is performed by a 1,100-actuator adaptive secondary mirror. In LTAO mode a 7.5\arcsec field of view is corrected to a level corresponding to a seeing-enhancement level a factor of a few better than the seeing and a factor of a few worse than the diffraction limit. Right: Typical (median conditions) MUSE NFM Point-Spread Function (PSF) with a Full-Width at Half Maximum of $\sim$ 70 mas at \halpha (656 nm) and a Moffat 1D profile. An overlay of a cartoon of the PDS 70 b/c system has been added to show the power of ELSDI, the planets being located in the photon-noise limited halo when collapsed in wavelength.}
\label{fig:parameterspace}
\end{center}
\end{figure}

\subsection{Medium resolution spectroscopy and LTAO}

Medium-resolution IFS at (or close to) the diffraction limit are very powerful to reach decent contrasts, appropriate to detect protoplanets, even in the absence of a coronagraph. This is the case with Keck/OSIRIS\cite{konopacky2013}, Gemini/NIFS and VLT/SINFONI\cite{hoeijmakers2018} in the NIR and now in the visible with MUSE NFM. Spectral resolutions from $\sim$1,500 to $\sim$5,000 allow to cross-correlate a model spectral template to the data and "lock" on molecular features of the spectrum. This was achieved on the planet $\beta$ Pictoris b\cite{hoeijmakers2018}.

\begin{figure}[h!]
\begin{center}
\includegraphics[width=0.65\textwidth, angle=0]{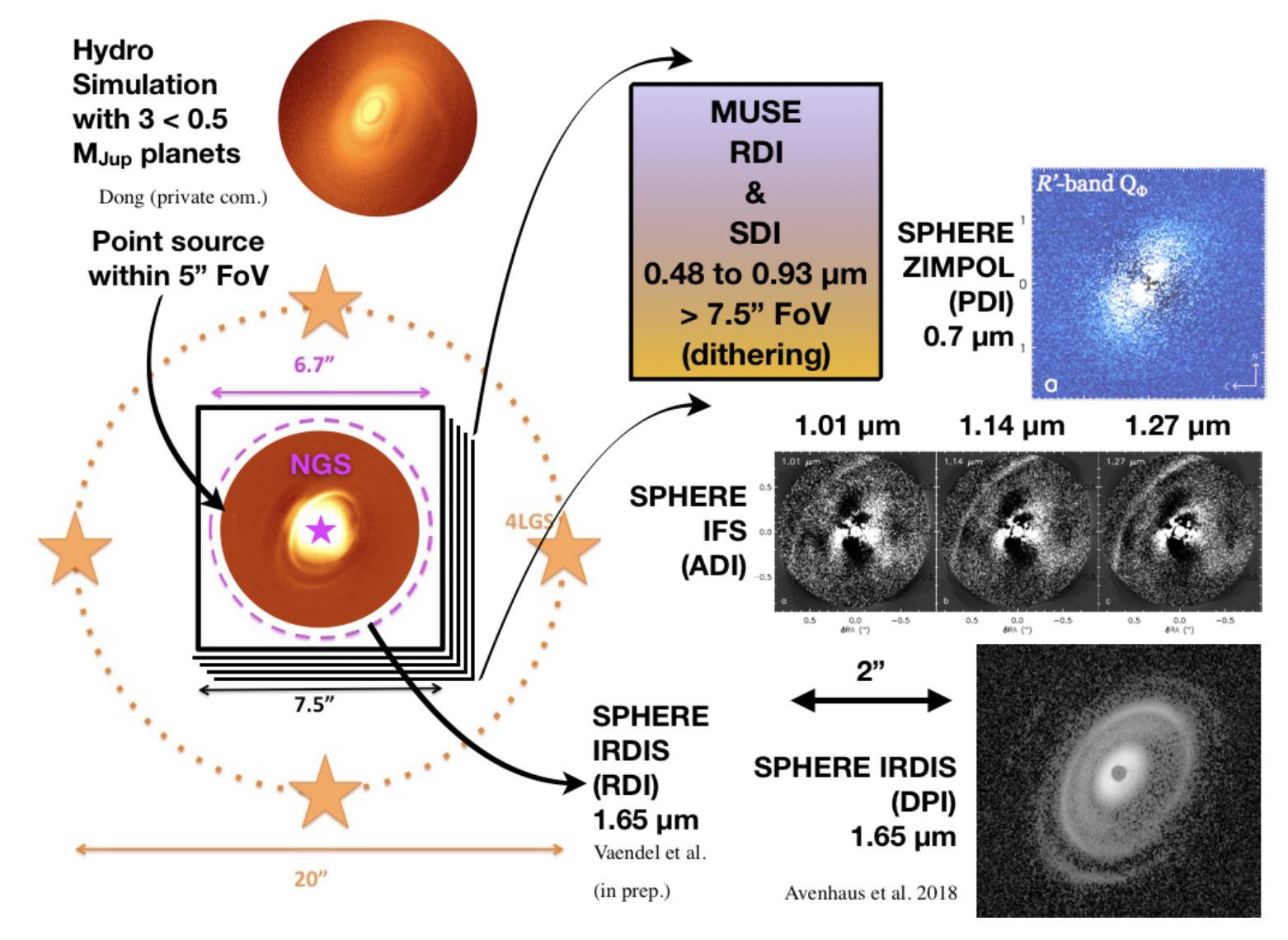}
\caption{Initial Science Verification idea to  attempt a panchromatic, high contrast imaging of RXJ1615 (top-right image in figure~\ref{fig:where}) with SPHERE, from 0.7 \micron (ZIMPOL) to 1.7 \micron (IRDIS). MUSE covers 0.48 to 0.93 \micron, At wavelength $\geq$ 0.8\micron we can expect to have a similar image quality than SPHERE?s IFS at 1\micron. On top/left is a hydrodynamic simulation to reproduce our target and its gaps by introducing 3 low-mass ($\leq$ 1 \mjup)?, yet giant planets (R. Dong, private com.). The data is being analyzed but it is unfortunately too shallow to achieve our goals.}
\label{fig:svgirard}
\end{center}
\end{figure}

Our team observed PDS 70 during the commissioning of the so-called MUSE narrow-field-mode (MUSE NFM) just prior to its Science Verification\cite{leibundgut2019} campaign. This mode and facility are unique in the world. They combine a novel Laser Tomographic Adaptive Optics (LTAO) system and a medium resolution Integral Field Spectrograph (IFS) covering the 0.4 to 0.9?m range with 25 milliarcseconds (mas) spaxels. The four-laser LTAO system, under good atmospheric conditions can routinely provide images with a full-width at half maximum (FWHM) of $\sim$50-100 mas at 600-700 nm over the 7.4\arcsec field of view (FoV). This is unprecedented and well adapted for T Tauri stars which are red and rather faint. Another advantage with MUSE is that the tip/tilt sensing is done in the NIR which is advantageous over most extreme AO systems that require bright natural guide stars (typically R or I $\leq$ 11).


\section{A second planet around PDS 70!}

In Haffert et al. 2019\cite{haffert2019} (Haffert19 hereafter) our team published the unambiguous detection of not one but two \halpha signals around PDS 70\cite{haffert2019}. One is at the location of planet b, confirming the presence of an accreting protoplanets and the other one is at the gap edge to the East and is most likely a second, accreting protoplanet. 

\begin{figure}[h!]
\begin{center}
\includegraphics[width=0.65\textwidth, angle=0]{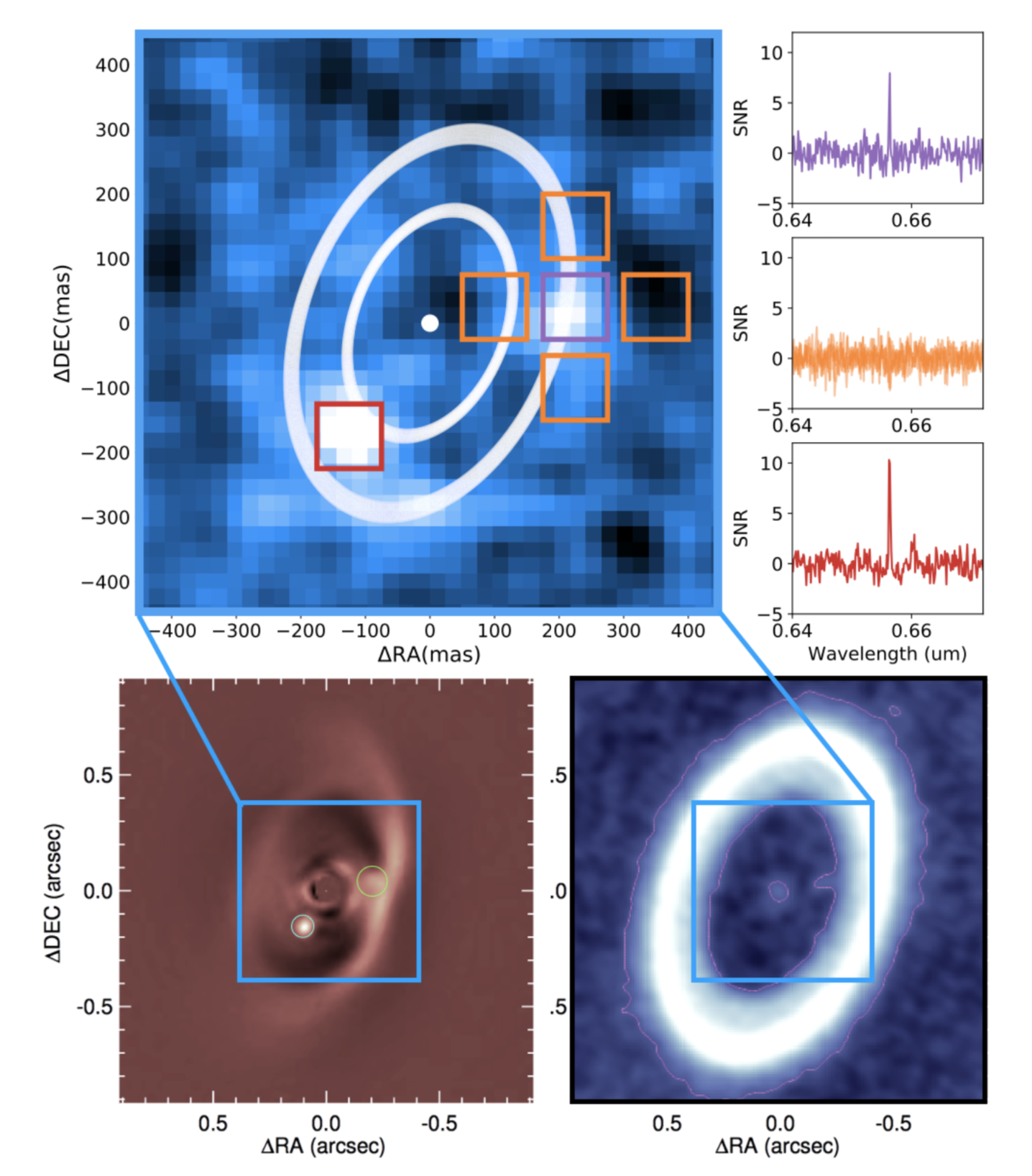}
\caption{Top: Our VLT/MUSE detection (\halpha SNR map) of both b and c \cite{haffert2019}, Press Release on June 3 2019) with to the right, the SNR versus wavelength around the \halpha line integrated in the corresponding color boxes (red for b and purple for c). The ellipse overlays show the orbital radii for both companions assuming circular Keplerian orbits. Bottom left: PDS 70 b as detected by VLT/SPHERE in K1K2\cite{muller2018}. Planet c is visible though not claimed at the time and called "bridge".Bottom right: ALMA Cycle 5 350.6 GHz continuum image with contours displaying a "spur" roughly at the location of c \cite{keppler2019}.}
\label{fig:pds70c}
\end{center}
\end{figure}

PDS 70 c is likely less massive, orbits in 2:1 mean motion resonance at the edge of the disk gap. Its location matches well a "spur" seen in the  ALMA continuum map (figure~\ref{fig:pds70c}). Its position corresponds also to a bright 
"blob" in emission in the high SNR SPHERE K-band image. If one were to deconvolve the disk image in M\"uller et al. 2018, the protoplanet c would probably be grazing the disk. Something worth noting is that a theoretician had already guessed the necessary existence of c (Bae, private com.) base on hydrodynamical simulations (figure~\ref{fig: bae3D}) before Haffert19 came out.


MUSE NFM's formidable sensitivity when these planets are accreting at a rate $\sim$ 10$^{-8}$ to 10$^{-7}$ \mjupy, which is likely for such massive and young systems.  Both detections are unambiguous: the ?\halpha signal cannot be stellar, since it is velocity shifted with respect to the line of sight. Also, major advantages of the mid-resolution IFS or emission line SDI technique (ELSDI) include:

\begini
\item The continuum is perfectly subtracted. Thanks to R$\sim$2500 spectral resolution, only \halpha is picked up against the continuum, whose gradient is accounted for. We estimate that we are at least a factor $\sim$5 times more efficient than the two other facilities that currently offer \halpha differential imaging through narrow (R$\sim$100) filters as the the expected contrast decreases quadratically as R increases: MagAO and VLT/SPHERE/ZIMPOL. As a confirmation to this statement, we have been able to detect again PDS 70 b and c in only 5 min (1 single 300-second cube) of MUSE NFM data while ZIMPOL has failed to reach these detection limits in $\sim$1h of data \cite{cugno2019}.
\item The accretion rate can be directly determined using the \halpha line width, instead of using a continuum contrast measurement like in the NIR, a method which is highly dependent on evolutionary models and the (uncertain) age of the star.
\item In the newly published high resolution ALMA data\cite{keppler2019, isella2019}, 3D hydrodynamic simulations\cite{bae2019} injecting a planet with a mass of 10 \mjup does not fully explain the disk morphology (wide gap). On-going calculations show (independently from the discovery announced in Haffert19) that a second, lighter planet in 2:1 mean motion resonance is a likely scenario and that?s precisely what our interpretation of our MUSE detection maps is. This is extremely exciting to have several teams come to similar conclusions using very different approaches and data.
\endi
Possible limitations to the ELSDI technique are the following:
\begini
\item Optical depth from circumstellar disk material?: planets have to be accessible to the line of sight (as for PDS 70 b) to detect \halpha emission. Some of the systems shown in Figure 2 still harbor gas (e.g., TW Hya, HL Tau, Elias 27, detecting \halpha could be challenging. No detections are also insightful: upper limits on the presence of companion(s) and constraints on the temperature of the gas, the size of the dust particles that scatter light away.
\item Accretion rate of the host star?: PDS 70's rate os  10$^{-11}$ \msuny. It can reach 10$^{-6}$ \msuny (10$^5$? more) for certain, younger star (age $\leq$10 Myr) and prevent detection of planets whose rate is of the order of 10$^{-8}$ to 10$^{-7}$ \mjupy (mind the units). In short, it?s trivial for weekly accreting stars. In that the detection space reported in figure~\ref{fig:parameterspace} is somewhat optimistic.
\endi

\begin{figure}[h!]
\begin{center}
\includegraphics[width=0.75\textwidth, angle=0]{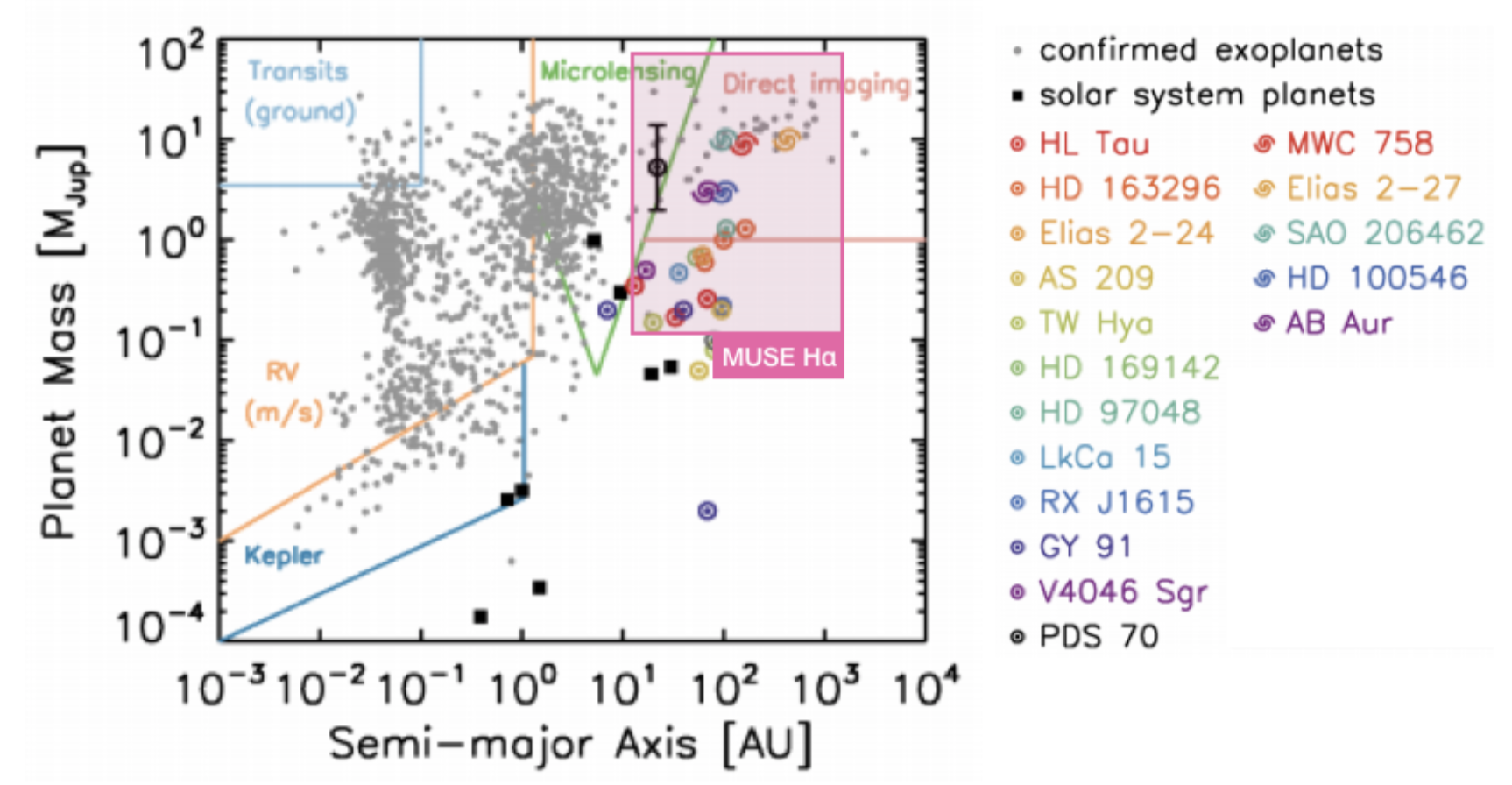}
\caption{The semi-major axis and mass of hypothesized planets assumed to reproduce the observed gaps in protoplanetary disks compiled from literature with an overlay of the parameter space that MUSE/NFM can probe (adapted from Bae {\sl et} al. 2018\cite{bae2018a}.}
\label{fig:parameterspace}
\end{center}
\end{figure}

\section{What are we looking at?}

A number of paper have already been published, in particular analyzing new ALMA data. Recently it was reported that  there is ALMA an signal at the location of b but not exactly, it is shifted by a few au\cite{isella2019}. Theoretical work also suggests that what can be imaged at NIR wavelengths are not the planet themselves (too embedded) but rather circumplanetatry disks (CPDs)\cite{szulagyi2019a}. 

\subsection{Hydrodynamic simulations}
While some theoretical modeling studies determined that a single planet was sufficient to carve the gap observed in PDS 70\cite{muley2019}, others showed that one planet was not enough (Fig\~ref{fig:bae3D}.

\begin{figure}[h!]
\begin{center}
\includegraphics[width=0.85\textwidth, angle=0]{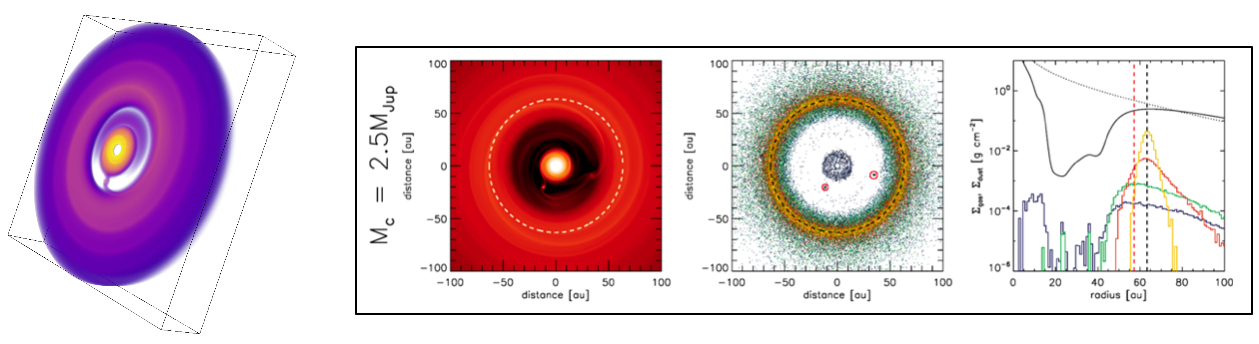}
\caption{Left: 3D hydrodynamic simulation of PDS 70 with only one planet (Bae, private com.) which cannot reproduce the observed gap (here a 5 \mjup planet). Right: PDS 70 b and c in mean motion resonance\cite{bae2019}. While the exact masses have yet to be determined these simulations reproduce the observations well (here, the mass of planet c $M_c$ = 2.5 \mjup, half that of planet b.}
\label{fig:bae3D}
\end{center}
\end{figure}

\subsection{Evidence of a circumplanetary disk around b}

Christiaens et al. published two papers based on VLT/SINFONI observations of PDS 70\cite{christiaens2019a, christiaens2019b}. They detect clearly b performing angular differential imaging (ADI) coupled with Spectral Differential Imaging (SDI). They fail to detect Br$\Gamma$ emission at the location of b or c. They also report somewhat complex features within the disk gap in thermal emission, which is surprising. 

Last bu not least, in the second 2019 paper, the presence of a CPD around b is revealed because of infrared excess with respect to theoretical models for such a planet. They report accretion compatible accretion rates both with previous measurements. For planet b, they derive an effective temperature of 1500?1600 K, a surface gravity log(g)$\sim$4.0, a radius of about 1.6. times that of Jupiter (R$_{Jup}$) , a mass around 10\mjup , and possible thick clouds. This all corroborates well with Haffert2019 and other studies.

\section{Conclusion \& future prospects}

In the future, plans are to extend even more the parameter space in angular resolution, contrast and sensitivity to go after objects that were not detectable so far, increasing the bound giant planet and brown dwarf demographics towards colder, fainter objects (i.e with JWST coronagraphy\cite{beichman2014_nircam, girard2018spie, perrin2018_JWST_coro_perfs}). For atmosphere characterization, DI will eventually become the method of choice for future flagship missions when contrasts of 10$^8$ to10$^9$ are routinely achievable (i.e when WFIRST CGI demonstrates wavefront control in space\cite{bailey2018}). The clear advantage over transit spectroscopy (the method of choice Today) is that the planet is always accessible, not just during transit. In parallel to pushing the limits in contrast through clever wavefront control hardware and algorithms, a somewhat new prospect arose: combine the detection capability of high dispersion spectroscopy (HDS) with high contrast imaging (HCI). There are several adaptations to existing facilities to do that (i.e KPIC at Keck\cite{wang2017HDS, mawet2017}, CRIRES$+$ and SPHERE at the VLT\cite{beuzit2019sphere, vigan2018hirise} and all the Extremely Large Telescope (ELT) programs in Europe and North America (ELT, GMT and TMT) have plans for instruments that will combine HDS and HCI\cite{riaud2007, snellen2015, lovis2017}. This is very much complementary to what can be done in space (greater stability, sensitivity, contrast but more constraints on weight and multiplexing to accommodate large and bulky spectrograph optics).\\

Meanwhile, MUSE NFM has an edge in that it can, with ELSDI reveal accreting protoplanets in \halpha (eventually other lines like \hbeta) with greater sensitivity than SPHERE/ZIMPOL or MagAO/Narrow-band SDI. One can regret though the absence of a focal plane coronagraphic mask in MUSE that would allow the studies of brighter targets with high efficiency (as for now, short exposure times and long, readout times are mandatory to avoid saturation around R $\leq$ 8 targets). PDS 70 b/c could be a nice "easy" target for WFIRST CGI low resolution spectroscopy mode with an \halpha filter, to measure, for instance, the variability of its accretion rates or simply to demonstrate its capability during the technology demonstration phase at relative low cost (modest contrast without fine tip/tilt and observing time).

\acknowledgments 
 
Julien Girard  thanks the Scientific Organizing Committee for the opportunity to speak (opening talk) at this great conference. We thank Ruobing Dong for useful discussion to prepare the SV program, ESO and its staff for operating and maintaining such a incredible facility and the MUSE consortium and all the technicians, scientists and engineers who have built such a unique instrument. P.Z. acknowledges support by the Forschungsstipendium (ZE 1159/1-1) of the German Research Foundation, particularly via the project 398719443.

\bibliography{AO4ELT2019} 
\bibliographystyle{spiebib} 

\end{document}